\documentclass[12pt]{article}
\usepackage{graphicx} 
\usepackage[utf8]{inputenc}
\usepackage[
backend=biber,
doi = true,
style=nature]{biblatex}

\usepackage{tikz}
\usetikzlibrary{calc}
\usetikzlibrary{positioning}
\usetikzlibrary{arrows}
\usepackage{comment}
\usepackage{amsmath}
\usepackage{amssymb}
\usepackage{subcaption}
\usepackage{mleftright}
\usepackage{hyperref}
\usepackage{bm}
\usepackage{bbm}
\usepackage{setspace}
\usepackage{romannum}
\usepackage[autostyle]{csquotes}
\usepackage{booktabs}
\usepackage{accents}
\usepackage{authblk}

\newcommand{\mc}[1]{\mathcal{#1}}

\usepackage[margin=2.5cm]{geometry}

\begin{document}
\pagenumbering{arabic}

\title{Short-term rental market occupancy - daily time series for $2017-2022$ on $500$ markets worldwide}
\author[1,2]{Marthe Elisabeth Aastveit}
\author[3]{Andreas Buscherm{\"o}hle}
\author[1,*]{Alex Lenkoski} 
\author[2]{Thordis Thorarinsdottir}

\affil[1]{Norwegian Computing Center, Oslo, Norway}
\affil[2]{Department of Mathematics, University of Oslo, Oslo, Norway}
\affil[3]{Wheelhouse, San Francisco, U.S.A.}
\affil[*]{Corresponding author: Alex Lenkoski: lenkoski@nr.no}

\date{\today}
\maketitle

\begin{abstract}
\noindent
Short-term vacation rentals, as promoted by platforms such as Airbnb, Homeaway, Vrbo, etc., are a growing component of the travel industry. This paper provides a  unique, large dataset on global market occupancy for the short-term rental market using data from the American company Wheelhouse. The dataset consists of data for $500$ markets around the world. For each market, a daily occupancy time series from January $2017$ to December $2022$ is provided, allowing for studies of local and global patterns in the evolution of the short-term rental market. Additionally, the dataset includes curves representing the booking trajectory of each market and stay date up to one year prior to the stay date. This large dataset comprises a unique combination of time series and survival analysis data, and is suitable as a methodological benchmark for both classical statistical and machine learning models.
\end{abstract}

\section{Background and Summary}
Short-term vacation rentals, commonly facilitated by platforms such as Airbnb, Homeaway and Vrbo, are the preferred accommodation type for many travelers, making them an important component of the travel industry. The travel industry constitutes a significant proportion of many countries' economies, and is particularly vulnerable to economic shocks. The shocks can both be on the supply and the demand side; the largest recent negative shock being the COVID-19 pandemic. The short-term rental market's response to this shock has been subject of extensive research \autocite{SaraDolnicar2021ABDa, lee2021consumers, kourtit2022airbnb, PAPPAS2021102767, BRESCIANI2021103170, PAPPAS2021104287}. It has been argued that vacation rentals have a higher level of resilience than the traditional hotel sector and therefore adapt faster to changes in demand patterns \autocite{li2019competitive}. The shock can then be viewed as smaller for the short-term rental market than for hotels \autocite{li2019competitive, medeiros2022exploring, SHARMA2021103180, GERWE2021120733}. 

Research topics of interest for the short-term rental market are highly diverse, from the effect Airbnb short-term rentals have on housing prices \autocite{Barron2021, Zhang2021, lee2016airbnb, horn2017home}, to psychology and customer behavior \autocite{TAJEDDINI2021102950, andreu2020airbnb, guttentag2018tourists}. More broadly, forecasting tourism demand has been a prominent research topic for both economists and statisticians alike, using classical time series models, econometric approaches or machine learning techniques\autocite{ATHANASOPOULOS200819, ATHANASOPOULOS2009146, ATHANASOPOULOS2011822, haensel2011booking, contessi2024decoding}. For example, Athanasopoulos {\em et al.} \autocite{ATHANASOPOULOS2011822} organised a tourism forecasting competition that compared a range of methods, including univariate and multivariate time series models as well as econometric approaches. Other methodological studies--e.g., Contessi {\em et al.} \autocite{contessi2024decoding} and Haensel {\em et al.} \autocite{haensel2011booking}--focused on individual hotels within specific markets. 

Currently available datasets on the short-term rental market commonly provide information at the host level for selected locations and stay dates\autocite{insideairbnb, tomslee, zadel2020influence}. AirDNA is a technology company that mines publicly available data from vacation rentals. AirDNA's data has been considered the best source for Airbnb statistics \autocite{zadel2020influence} and several studies have been based on these data \autocite{Gunther2025, medeiros2022exploring, zadel2020influence}. Inside Airbnb is a worldwide group of residents, activists and allies from different organizations\autocite{insideairbnb} who provide data for studying the effect Airbnb has on residential communities. Inside Airbnb mainly provides information at the host level, on specific hosts and room types, etc. While these data are widely used in academic research, it has been suggested there may be issues with systemic errors in the data collection process\autocite{ALSUDAIS2021113453}. Additionally, individuals have collected and published datasets. For example, from $2013-2017$, Tom Slee gathered host-level data on Airbnb rentals for several markets around the world\autocite{tomslee}. 

In contrast, the current study provides market-level data on short-term rental market occupancy for $500$ markets around the world for the time period January $1$st, $2017$, to December $31$st, $2022$. The data have been collected by the American company Wheelhouse who provide a property management and dynamic pricing platform for short-term vacation rental markets worldwide\autocite{Wheelhouse}. These unique aggregated time series data provide daily information on market occupancy, the estimated number of bookings and the estimated number of available units. The dataset furthermore includes booking curves representing the daily booking trajectory of each market and stay date up to one year prior to the stay date. To our knowledge, this is the first publicly available dataset of market occupancy curves for the short-term rental market, opening up new opportunities for modeling market-level behaviors. 

The objective of releasing this dataset is two-fold. First, the data provide a unique insight into the short-term rental market and may be used, on their own or along with other data sources, to gain insights into market behaviors. Potential open research questions include occupancy patterns during the COVID-19 pandemic, changes in the global market occupancy over time and correlation patterns between market occupancy and house prices. Secondly, the data -- with their intricate space-time structure -- present a unique large and complex dataset that may be useful for general methodological research in machine learning, statistics and econometrics, especially when benchmarking novel forecasting methodologies.  

\section{Methods}
\subsection{Data collection}\label{subsec:data coll}
The raw data underlying the current dataset were collected through a variety of sources, including monitoring short-term vacation rental websites and directly linked information. The most recent calendar data are collected for a collection of units once per day. Comparing these data across dates provides information on how individual stay dates turn from available to unavailable or, potentially, become available again. A stay date may become unavailable for two reasons. The stay date either became booked by a guest, or the host blocked off the stay date, e.g. due to maintenance work or the owner using the property for themselves. To estimate actual guest occupancy, it is important to distinguish these two cases. For this, we use a proprietary classification algorithm that assigns a booking probability to each stay date. Through this daily monitoring of which stay dates became unavailable and assessment of their probability to have been booked, we obtain the raw data necessary to build up the aggregated dataset. 

The data have been collected over a wide time frame from 2016 through 2022 and daily operations on a large number of listings on Airbnb are bound to not always succeed without issues. Occasionally, data for individual listings or on a larger scale cannot be collected on a given day or for multiple days in a row. As a consequence, the time when a booking occurred becomes uncertain. This impacts the published dataset such that multiple stay dates in a row might see fewer booking events/increases in occupancy than actually happened, and then in a single day all these bookings are caught up, and the occupancy jumps to a higher level. We have chosen to leave these issues in the published data as they represent the best real information we have on the actual market behavior and can be used for anomaly detection or data imputation methodologies.

\subsection{Aggregating the data} \label{subsec:agg_data}
The daily market occupancy curve provided as a part of the dataset is the inverse of the survival curve. Specifically, for a survival function $\mathbb{S}(t)$ the market occupancy is $\mathbb{M}(t) = 1 - \mathbb{S}(t)$, where $t$ indicates time. Since the relevant theoretical background comes from survival analysis, we focus on describing the survival function $\mathbb{S}(t)$ rather than the market occupancy $\mathbb{M}(t)$. We first present the relevant theoretical foundations, before we connect these with the interpretation for the short-term rental market. 

The survival function $\mathbb{S}(t)$ is the expected proportion of individuals where an event has not happened by time $t$\autocite{aalen2008survival}. It is extensively used in medical research where the event commonly is death or the onset of a disease. Following Aalen {\em et al.}\autocite{aalen2008survival}, the survival function is formally written 
\begin{equation}
    \mathbb{S}(t) = Pr(T > t),
\end{equation}
where the random variable $T$ is the survival time. As an increasing proportion of individuals experience the event, the survival function will often tend towards $0$ as $t$ increases. However, the event need not occur for all individuals; consequently, the survival time $T$ may be infinite for some subjects and the survival function $\mathbb{S}(t)$ may converge to a positive limiting value\autocite{aalen2008survival}. 

Several approaches exist for estimating survival curves; for independent, right‑censored observations the Kaplan–Meier estimator remains the standard nonparametric estimator \autocite{kaplan1958nonparametric}. The independent (noninformative) right‑censoring assumption requires that, given survival up to $t$, the hazard of failure on $[t,t+\mathrm{d}t)$ is the same for individuals who will be censored shortly thereafter as for those who will not be censored, i.e. censoring carries no additional information about the failure time \autocite{aalen2008survival}. These assumptions are plausible for our data, and we therefore use the Kaplan–Meier estimator to estimate the survival functions.

Let $\mc{S}$ be a family of survival curves and define $\mathbb{S}$ to be a monotonically decreasing function, where $\mathbb{S}(t) \leq \mathbb{S}(r)$ for all $t > r$. Following general theory of Kaplan-Meier estimates\autocite{kaplan1958nonparametric, aalen2008survival}, we can write any $\mathbb{S}_i \in \mc{S}$ as
\begin{align}
        \mathbb{S}_i(t) & = \prod_{r\leq t} \Big( 1 - \frac{d_{ir}}{n_{ir}} \Big) \\
        & = \prod_{r\leq t}\xi_{ir},
    \label{eq:surv_1}
\end{align}
where $d_{ir}$ is the number of events that occurred at time $r$ and $n_{ir}$ is the number of individuals that are known to have survived up until time point $r$. This means $\xi_{ir} \in (0,1]$, and the factors $\xi_{ir}$ are the probabilities of surviving past time $r$, conditional on having survived up to time $r$\autocite{aalen2008survival}. 

As far as we are aware, we are the first to introduce survival analysis methods into the study of short‑term rental market data. The interpretation of the survival curve $\mathbb{S}(t)$ for these data is the expected proportion of vacation listings that have not yet been booked by time $t$. $d_{ir}$ is the estimated number of bookings that happened at time $r$, and $n_{ir}$ is the estimated number of listings that have survived up to time $r$. $d_{ir}$ and $n_{ir}$ are estimated, meaning we do not always know whether a listing is booked or blocked, as we discussed in Section \ref{subsec:data coll}. Note that the survival curve starts $365$ days before the stay date. That is, if we for example look at the stay date October $2$nd $2022$, the survival curve starts on October $2$nd $2021$. 

The baseline curve shows the survival curve path of a common stay date for a given market. Assume there exists a reference member $\mathbb{S}_0 \in \mc{S}$, which is the baseline survival function. The baseline survival function $\mathbb{S}_0$ can be defined in the same way as Equation \eqref{eq:surv_1}:
\begin{align}
    \mathbb{S}_0(t) &= \prod_{r\leq t} \Big( 1 - \frac{d_{0r}}{n_{0r}} \Big) \\
    & = \prod_{r\leq t}\xi_{0r},
    \label{eq:surv_0}
\end{align}
where $d_{0r}$ and $n_{0r}$ is the number of events and number at risk for all the stay dates. This is more formally written as
\begin{align}
    d_{0r} & = \sum_{i = 1}^{m} d_{ir} \\
    n_{0r} & = \sum_{i = 1}^{m} n_{ir},
\end{align}
where $m$ is the number of available stay dates for a given market. 

\subsection{Preserving data privacy}
The published dataset poses no data‑privacy concerns since it is provided in aggregated form; the underlying individual‑level raw data will not be released. By choosing large markets with hundreds to thousands of properties, it is impossible to infer individual-level information based on these data. 

\section{Data Records}
The data are separated in two tar compressed files. The first compressed file, called \textit{surv\_data\_2022.tar}, consists of one file with market information and $500$ .csv-files with the survival data. Each of the $500$ files represents one market. The second compressed file, called \textit{events\_risk\_data\_2022.tar}, consists of $500$ .csv-files with the estimated number of events (one file for each market) and $500$ .csv-files with the estimated number at risk (one file for each market). The data in the second compressed file (\textit{events\_risk\_data\_2022.tar}) can be used to calculate the data in the first compressed file (\textit{surv\_data\_2022.tar}). Relevant code is available on \url{https://github.com/NorskRegnesentral/ForecastMarketOcc/blob/master/DataPaper/making_surv_data.R}. 

We start with the first compressed file \textit{surv\_data\_2022.tar} called \textit{active\_markets.csv} which presents the connection between market number and market name. This .csv-file has four columns; market number, market name, state and country. State is only provided for the markets in the US and some markets in Canada. An illustration of the data table is presented in Table \ref{tab:market_names}.
\begin{center}
\captionof{table}{Examples of some market numbers, names, state and country.}\label{tab:market_names}
\begin{tabular}{p{2cm} p{4cm} p{2cm} p{3cm}} 
 \toprule
 \textbf{Market number} & \textbf{Market name} & \textbf{State} & \textbf{Country}\\ 
 \toprule
 $1$ & San Francisco & CA & USA\\ 
  \addlinespace[0.5em]
 $2$ & Lake Tahoe & CA & USA \\
 \addlinespace[0.5em]
 $3$ & Seattle & WA & USA \\
 \addlinespace[0.5em]
 \dots & \dots & \dots & \dots \\ 
  \addlinespace[0.5em]
 $112$ & Paris & & France\\ 
  \addlinespace[0.5em]
 $113$ & Barcelona & & Spain\\
  \addlinespace[0.5em]
 $114$ & Naples & & Italy\\
  \addlinespace[0.5em]
 \dots & \dots & \dots & \dots \\
  \addlinespace[0.5em]
 $193$ & Tokyo & & Japan \\
  \addlinespace[0.5em]
 $194$ & Luton & & UK \\
  \addlinespace[0.5em]
 $195$ & Hong Kong & & China\\
  \addlinespace[0.5em]
 \dots & \dots & \dots & \dots \\
  \addlinespace[0.5em]
 $500$ & Corfu & & Greece\\
 \bottomrule
\end{tabular}
\end{center}

For the survival data, the .csv-file for each market is called \textit{surv\_market\_j}, where \textit{j} is the market number. For example, market $1$ has the name \textit{surv\_market\_1.csv}. Each .csv-file consists of $367$ columns and the number of rows is equal to the number of stay dates available for this market. Specifically, each data file is a matrix of size $(m+1) \times 367$, where $m$ is the number of stay dates. The stay date ranges from January $1$st $2017$ to December $31$st $2022$. Note that for some markets, the first observed date is after January $1$st $2017$. 

The first row consists of the data for $\mathbb{S}_0$. Column $1$ consist of the stay dates. For row $1$ ($\mathbb{S}_0$) and column $1$, the value is $0$. Columns $2-367$ consists of data from $365$ days before the stay date to the actual date. For example, column $2$ consists of $\mathbb{S}_i(1)$, while column $367$ is $\mathbb{S}_i(366)$. Table \ref{tab:description} gives a description of the rows and columns in the dataset. 

\begin{center}
\captionof{table}{The format, with a description, of the rows and columns for the survival data. The data is stored as a separate .csv-file per a market.}\label{tab:description}
\begin{tabular}{p{2cm} p{3cm} p{10cm}} 
 \toprule
 \textbf{Row} & \textbf{Column} & \textbf{Description}\\ 
 \toprule
 \addlinespace[0.5em]
 $1$ & $1$ & Value of $0$ since this is the row for the baseline. \\ 
 \addlinespace[0.5em]
 $1$ & $2-367$ & $\mathbb{S}_0(t)$ (baseline) for $365$ days before the stay date (column $2$) to the actual stay date (column $367$).\\
 \addlinespace[0.5em]
 $2-m$ & $1$ & Stay dates. \\
 \addlinespace[0.5em]
 $2-m$ & $2-367$ & $\mathbb{S}(t)$ for $365$ days before the stay date (column $2$) to the actual stay date (column $367$).\\
 \bottomrule
\end{tabular}
\end{center}
Note that the dataset consists of the survival curves, and not the market occupancy curves. To study the market occupancy, calculate $\mathbb{M}(t) = 1 - \mathbb{S}(t)$.

The second compressed file consists of the number of events $d_{ir}$ and the number at risk $n_{ir}$ as was discussed in Section \ref{subsec:agg_data}. The events and risk files are separated in two different .csv-files per market. The name of the file for the number of events is \textit{n\_events\_market\_1.csv} for market 1, \textit{n\_events\_market\_2.csv} for market 2 and so on. For the number at risk, the name is \textit{n\_risk\_market\_1.csv} for market 1, \textit{n\_risk\_market\_2.csv} for market 2 and so on. Each .csv-file consists of $367$ columns and the number of rows are the number of stay dates. The format with rows and columns are equal in the files for events and at risk. 

\begin{center}
\captionof{table}{Description of the rows and columns in the events/risk .csv-files.}\label{tab:description_risk_events}
\begin{tabular}{p{2cm} p{3cm} p{10cm}} 
 \toprule
 \textbf{Row} & \textbf{Column} & \textbf{Description}\\ 
 \toprule
 $1-m$ & $1$ & Stay dates. \\ 
  \addlinespace[0.5em]
 $1-m$ & $2-367$ & Data for $d_{it}$ or $n_{it}$ for $365$ days before the stay date (column $2$) to the actual stay date (column $367$),\\
 \bottomrule
\end{tabular}
\end{center}

\section{Technical validation}\label{sec:tec}
In this section, we show market occupancy instead of survival curves for ease of exposition.

\subsection{The markets}
\begin{figure}[!htb]
\centering
        \includegraphics[page = 1, height=0.6\textwidth, width=\linewidth]{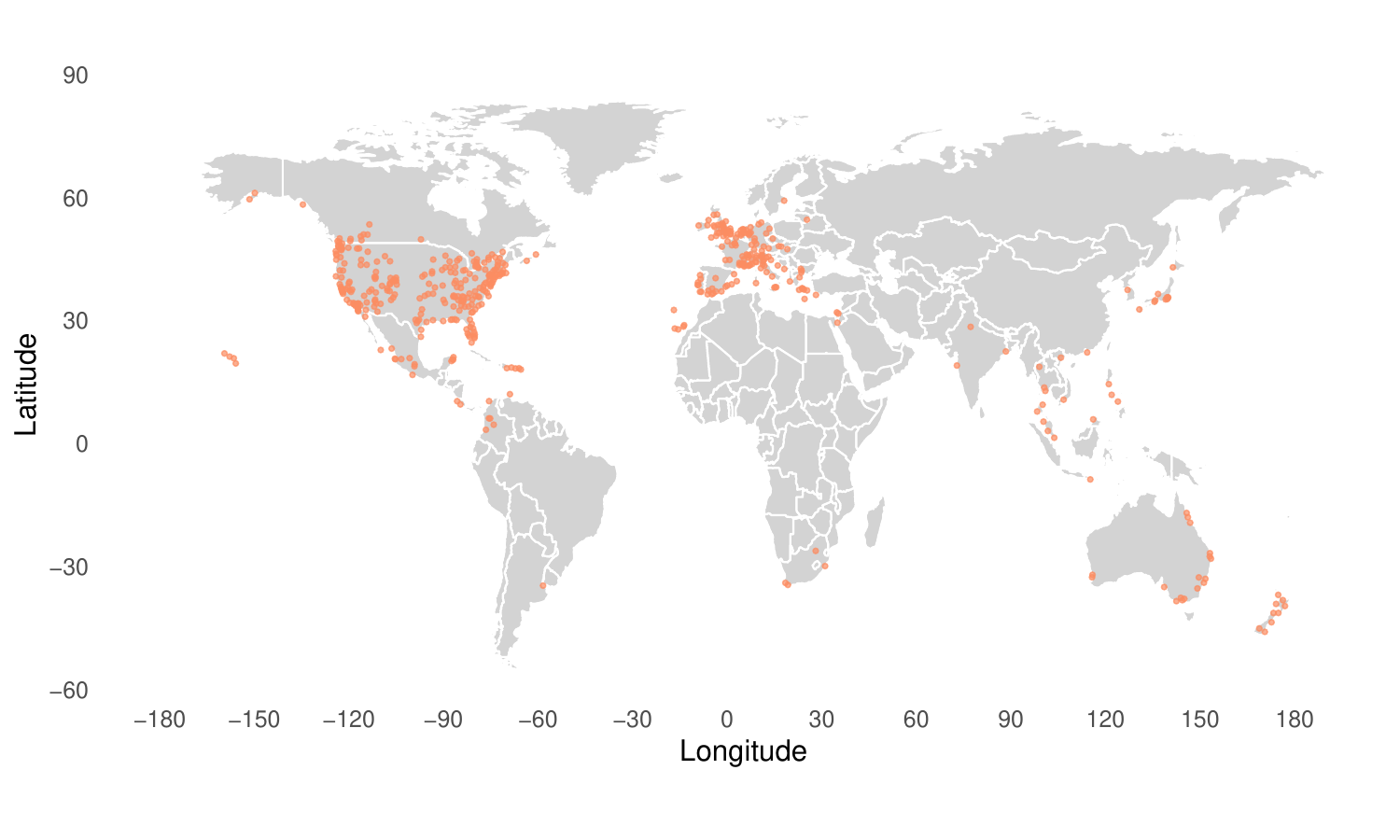}
        \caption{A map of all $500$ markets.}
        \label{fig:map}
\end{figure}
Figure \ref{fig:map} displays a map of all $500$ markets. The main part of the markets are located in the US and Europe, but there are markets covering the globe. The $500$ markets are a unique combination of different types of markets, from city-markets, like New York and San Francisco, to vacation destinations like Destin, FL, and Lake Tahoe, enabling a diverse set of studies to be conducted.    

\subsection{Baseline market occupancy}
\begin{figure}[!htb]
\centering
        \includegraphics[page = 1, height=0.5\textwidth, width=0.8\linewidth]{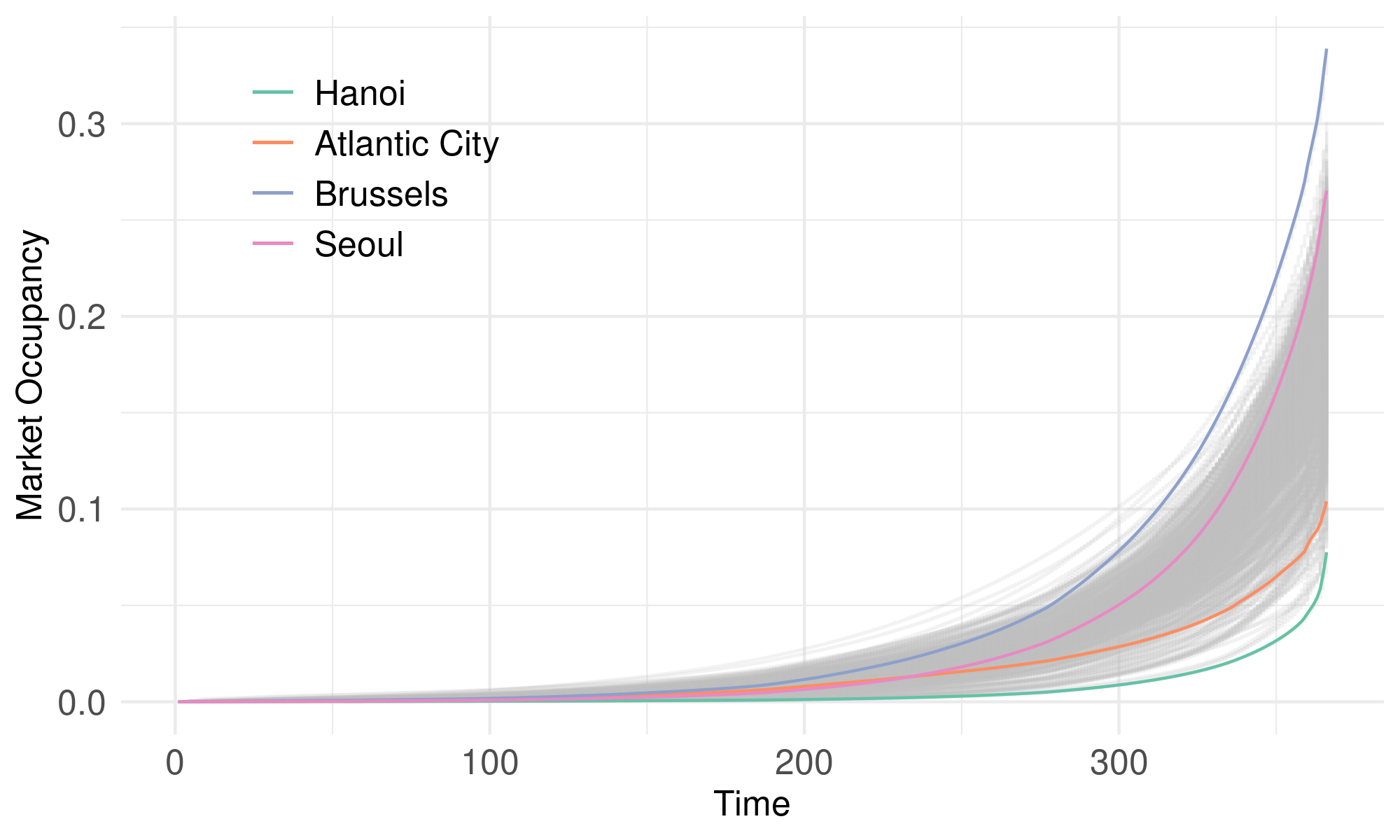}
        \caption{Baseline market occupancy for all the $500$ markets, where four markets (Hanoi, Atlantic City, Brussels and Seoul) are highlighted.}
        \label{fig:baseline_occ}
\end{figure}

In Figure \ref{fig:baseline_occ}, we show the baseline market occupancy ($\mathbb{M}_0(t) = 1 - \mathbb{S}_0(t))$ for all $500$ markets. We have highlighted two markets which end up with a lower occupancy, and two with a higher occupancy than normal. When $t = 1$, it means $365$ days before the stay date, more specifically $t = 366-k$, where $k \in [1, \dots, 366]$ is the number of days before the stay date. When $k = 366$, we report the final market occupancy. 

The baseline market occupancy curves indicate the proportion of the market which would be estimated to be occupied on a ``normal'' day by lead-time, without other information.  It can therefore serve as a ``mean'' model in e.g. a random effects framework as discussed in Aastveit {\em et al.} \autocite{Marthephd1paper}, as well as providing high level context regarding the market's behavior.  We have provided an overall estimate of this baseline rate, which uses data from all stay dates. This feature can also be computed on-the-fly from a subset of the stay dates by using the associated risk and events terms for these dates, as discussed in Section \ref{subsec:agg_data}.

\subsection{Example: Lake Tahoe and San Francisco}
\begin{figure}[!htb]
\centering
        \includegraphics[page = 1, height=0.5\textwidth, width=0.8\linewidth]{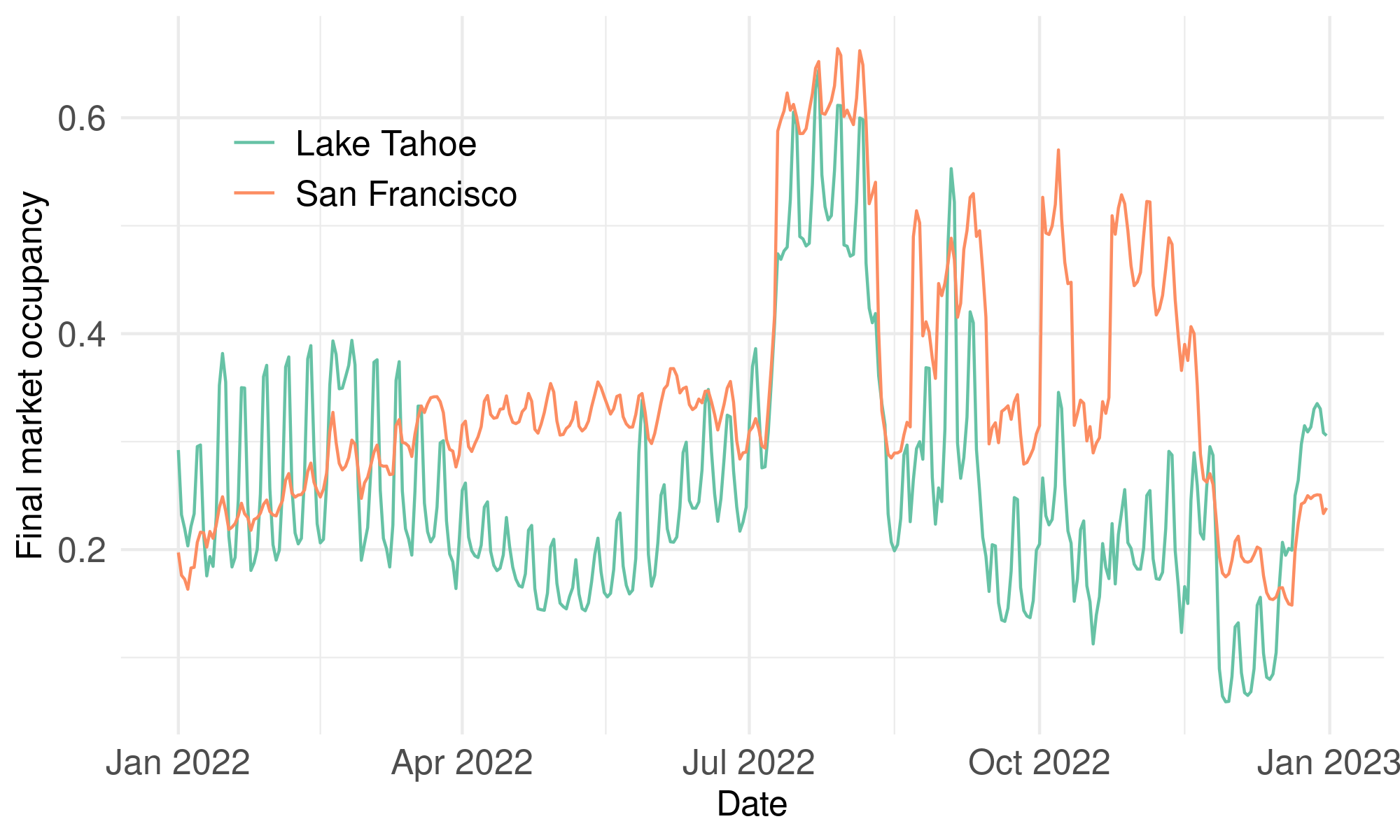}
        \caption{Final market occupancy in San Francisco and Lake Tahoe in $2022$.}
        \label{fig:final_occ}
\end{figure}

We illustrate the data by studying two example markets; San Francisco and Lake Tahoe. San Francisco is a typical city market with weekend effects and events throughout the year. As a vacation destination, Lake Tahoe is popular both during winter (skiing) and summer (lake activities). The final market occupancy for $2022$ is illustrated in Figure \ref{fig:final_occ}. In this plot, the seasonality for Lake Tahoe is clear, with an overall rise in the occupancy in summer and some rise during the winter. For San Francisco, the final market occupancy is quite stable during the winter and spring. There is a sudden jump in the summer, and the final market occupancy continues to be high throughout the fall. This can be a consequence of the COVID-19 pandemic.   

\begin{figure}[!htb]
    \begin{subfigure}{0.5\linewidth}
        \includegraphics[page = 1, height=0.3\textheight, width=\linewidth]{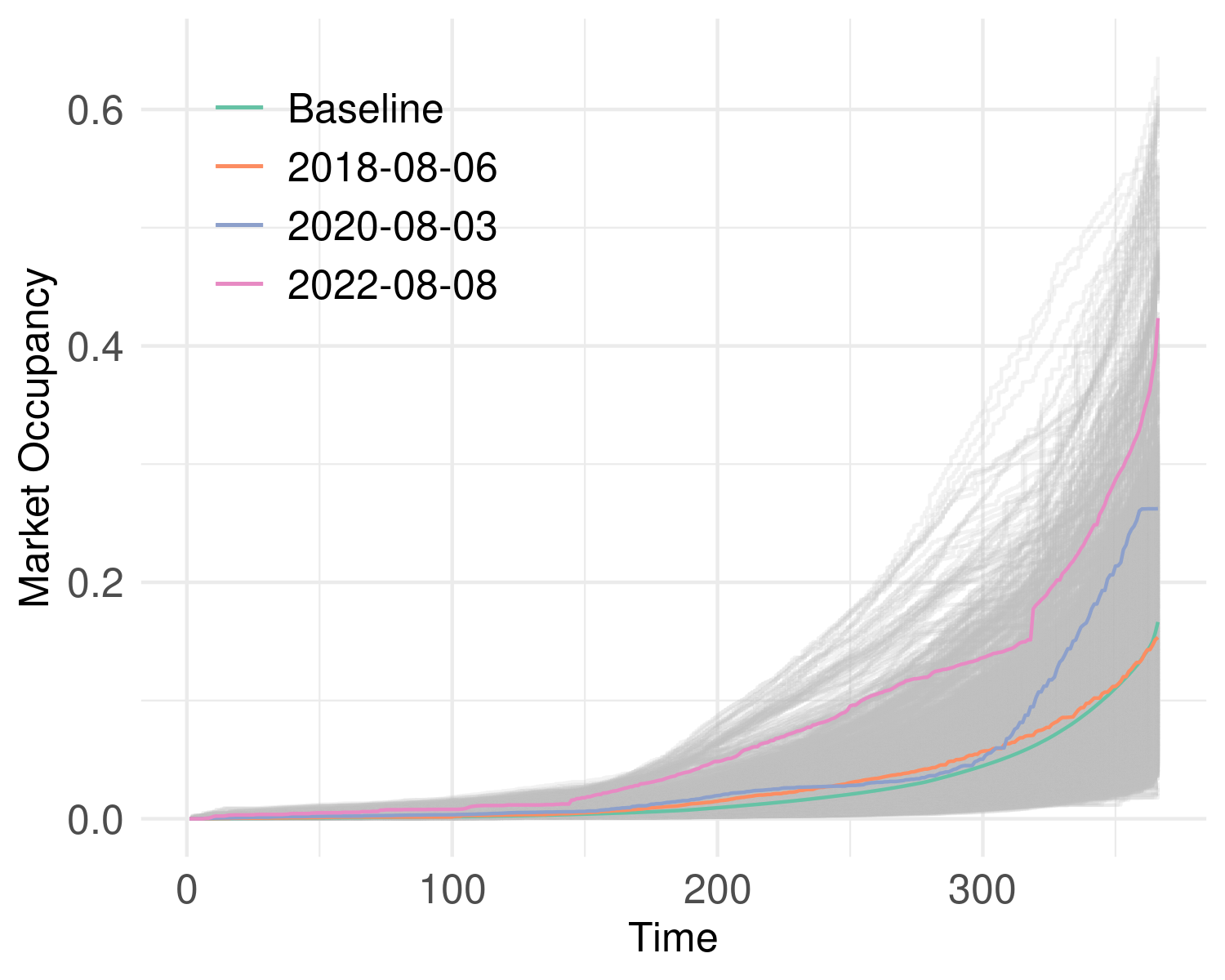}
        \caption{$M_{i}(t)$ for Lake Tahoe.}
        \label{fig:occ_LT}
    \end{subfigure}
    \hfill
    \begin{subfigure}{0.5\linewidth}
        \includegraphics[page = 1, height=0.3\textheight, width=\linewidth]{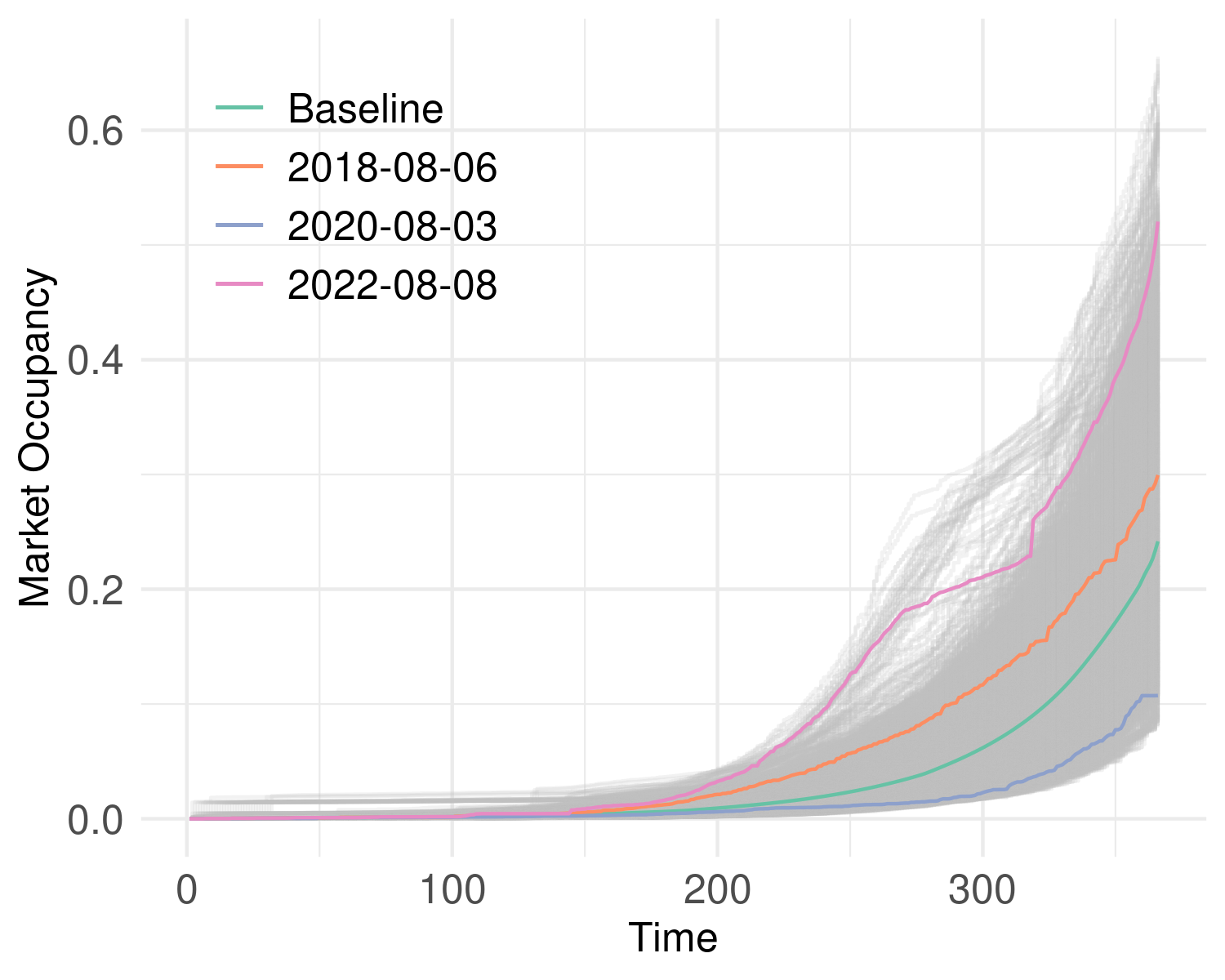}
        \caption{$M_{i}(t)$ for San Francisco.}
        \label{fig:occ_SF}
    \end{subfigure}
    \caption{$M_{i}(t)$ for Lake Tahoe and San Francisco. The gray lines are $M_{i}(t)$ for all the stay dates in each market. Three Mondays in the beginning of August in $2018$ (orange), $2020$ (blue) and $2022$ (pink) are highlighted, in addition to the baseline.}
    \label{fig:occ_SF_LT}
\end{figure}

Figures \ref{fig:occ_LT} and \ref{fig:occ_SF} show the market occupancy paths for Lake Tahoe and San Francisco. Each gray line represents the market occupancy trajectory for one specific stay date, and all the gray lines together represents all the stay dates in the data. The baseline and three stay dates are highlighted. The stay dates are a Monday in beginning of August in $2018$, $2020$ and $2022$, to illustrate a path before, during and right after the COVID-19 pandemic. Figures \ref{fig:occ_LT} and \ref{fig:occ_SF} then illustrate how these markets behaved differently during the pandemic. Lake Tahoe was a typical national travel destination during the pandemic, with space for hiking and swimming. The blue line in Figure \ref{fig:occ_LT} was low and stable until around $60$ days before the stay date. Then people dared to book their summer vacation only two months prior to traveling, and consequently the occupancy increased fast. San Francisco however, was mostly closed during the pandemic. This is illustrated with the blue line in Figure \ref{fig:occ_SF}, which is always low.

Other things to note in Figures \ref{fig:occ_LT} and \ref{fig:occ_SF} is a clear jump around $\text{time} = 300$ in the line for August $8$th $2022$ in both markets. The reason is database issues in the raw data. Suddenly more listings were marked as booked, but the listing was for example booked the previous day. This was discussed in Section \ref{subsec:data coll}. We chose to leave these jumps in the dataset, allowing researchers to develop imputation and fault detection methods on these data.

\section{Data availability}
The host of the data is dataverse at University of Oslo (UiO). The dataset is currently not publicly available, but will be deposited and made available upon publication of the associated article. This manuscript will be updated with the persistent dataset DOI/link when it is released.

The data are in two tar compressed files. In surv\_data\_2022.tar, there are $500$ .csv of the survival data, one for each market. In addition, there is a .csv-file called active\_markets.csv, where the information about market number, market name, state and country is provided. The second compressed file is called events\_risk\_data\_2022.tar, and consists fo data with number of events ($500$ .csv-files) and number at risk ($500$ .csv-files).

\section{Code availability}
The raw data for estimating number of events and number at risk are not openly available, and will not be published. Relevant code is found on Github \url{https://github.com/NorskRegnesentral/ForecastMarketOcc/DataPaper}. In the Github package, there is a folder called DataPaper, where the code for making the plots and generating $\mathbb{S}(t)$ from $n_{ir}$ (number at risk) and $d_{ir}$ (number of events).  

\clearpage
\printbibliography 

\section{Author Contributions}
Contribution:\\
Marthe Elisabeth Aastveit: Data curation, Writing - Original draft preparation, Methodology, Coding. \\
Andreas Buscherm{\"o}hle: Data curation, Writing (Data Collection).\\
Alex Lenkoski: Data curation, Writing, Methodology, Coding.\\
Thordis Thorarinsdottir: Writing. \\
All authors reviewed and approved the final manuscript.

\section{Competing Interests}
The authors declare no competing interests. 

\section{Acknowledgments}
MEA acknowledges the support of the Research Council of Norway through grant 342613 "Predicting in High Dimensions". TLT was supported by the Research Council of Norway through the Centre of Excellence "Integreat – The Norwegian Centre for Knowledge-driven Machine Learning", project number 332645. 

\section{Funding}
This work was not funded by any external source.

\end{document}